%% file: avr.tex
\documentclass[useAMS,usenatbib]{mnras}

\usepackage{amsmath,amssymb,amsfonts}
\usepackage{bm}
\usepackage{graphicx}
\usepackage{color}
\usepackage[varg]{txfonts}
\usepackage{enumerate}
\usepackage{enumitem}

\newcommand{\kms}{km\,s$^{-1}$}
\newcommand{\z}{$|z|$}
\newcommand{\sigR}{$\sigma_{\rm R}$}
\newcommand{\sigPhi}{$\sigma_\phi$}
\newcommand{\sigZ}{$\sigma_{\rm z}$}
\newcommand{\feh}{${\rm [Fe/H]}$}
\newcommand{\afe}{${\rm [\alpha/Fe]}$}

\begin{document}

\title[The AVR of the Galactic discs]{The age-velocity dispersion relation of the Galactic discs from LAMOST-\emph{Gaia} data}

\author[Jincheng~Yu \& Chao~Liu]{Jincheng~Yu\thanks{email: yujc.astro@gmail.com}
\& Chao~Liu \\
National Astronomical Observatories, Chinese Academy of Sciences, Beijing 100012, China \\
}

\date{Manuscript Version: Aug 2017}

\pagerange{\pageref{firstpage}--\pageref{lastpage}} \pubyear{2017}

\maketitle

\label{firstpage}

\begin{abstract}
	We present the age-velocity dispersion relation (AVR) in three dimensions in the solar neighbourhood using 3,564 commonly observed sub-giant/red-giant branch stars selected from LAMOST, which gives the age and radial velocity, and \emph{Gaia}, which measures the distance and proper motion. The stars are separated into metal-poor (${\rm [Fe/H]<-0.2}$\,dex and metal-rich (${\rm [Fe/H]>-0.2}$\,dex) groups, so that the metal-rich stars are mostly $\alpha$-poor, while the metal-poor group are mostly contributed by $\alpha$-enhanced stars. Thus, the old and metal-poor stars likely belong to the chemically defined thick disc population, while the metal-rich sample is dominated by the thin disc. The AVR for the metal-poor sample shows an abrupt increase at $\gtrsim7$\,Gyr, which is contributed by the thick disc component. On the other hand, most of the thin disc stars with ${\rm [Fe/H]>-0.2}$\,dex display a power-law like AVR with indices of about 0.3--0.4 and 0.5 for the in-plane and vertical dispersions, respectively. This is consistent with the scenario that the disc is gradually heated by the spiral arms and/or the giant molecular clouds. Moreover, the older thin disc stars ($>7$\,Gyr) have a rounder velocity ellipsoid, i.e.~$\sigma_\phi/\sigma_{\rm z}$ is close to 1.0, probably due to the more efficient heating in vertical direction. Particularly for the old metal-poor sample located with $|z|>270$\,pc, the vertical dispersion is even larger than its azimuthal counterpart. Finally, the vertex deviations and the tilt angles are plausibly around zero with large uncertainties.
\end{abstract}

\begin{keywords}
	Galaxy: evolution -- Galaxy: disc -- Galaxy: kinematics and dynamics -- solar neighbourhood
\end{keywords}

\section{Introduction}
\label{sec:introduction}

The age-kinematics relation of the stars in the solar neighbourhood provides crucial information on the structure and evolution of the Milky Way. The velocity ellipsoid, as a descriptive quantity of the distribution function, for these stars reflects the gravitational potential and hence the spatial structure, while the varying kinematics with the stellar ages can constrain the evolution of the Galaxy.

Older stars are observed to have larger velocity dispersions \citep{casa2011}, while younger stars are typically kinematically cool \citep{aume2009}. The increasing trend of the velocity dispersions with age is known as the age-velocity dispersion relation (AVR) and has been discussed for decades \citep[e.g.][]{stro1946, wiel1977, nord2004}. \cite{quil2001} detected a jump in the AVR for stars older than 9\,Gyr from a sample of F and G dwarf stars and identified the jump as a thick disc component caused by minor merger. Other works favours a simple power law \citep*[][]{aume2009, holm2009} with a saturation for stars older than 8\,Gyr.

One possible explanation is that the stars are born with a decreasing velocity dispersion, since the turbulence of gas discs decreases with age due to the decreasing gas accretion rates \citep*{bour2009, forb2012, bird2013}. \cite{ma2017} found that stars older than 6\,Gyr were formed in a violent mode, while stars younger than 6 Gyr\,were formed in a relatively calm disc in a cosmological zoom-in simulation. This view is also supported by observations of gas kinematics in disc galaxies which shows a decline of intrinsic velocity dispersion with decreasing redshift \citep{wisn2015}. However, the observed clumpy galaxies might be too massive to be the progenitors of Milky Way-type galaxies \citep[e.g.][]{vdok2013, inou2014}. Furthermore, it is unclear how the kinematics of young stars which form from cold gas relate to the observations.

Most theoretical works treated the observed AVR as a consequence of gradual heating through scattering processes, with the assumption that stars are born with a roughly constant velocity dispersion probably smaller than 10\,\kms\ \citep[][]{aume2009}. Both massive gas clouds \citep{spit1953} and spiral arms \citep{barb1967} are considered as scattering agents.

\cite{sell1984} and \cite{carl1985} showed that the transient spirals can significantly heat the in-plane motion of disc stars without much influence the vertical motion \citep[e.g.][]{sell2013,mart2015}. However, the observed \sigZ\ increases similarly with age as the in-plane components \citep{holm2009}, implying that the spiral arms cannot be the only source.

\cite{spit1953} showed that giant molecular clouds (GMCs) with masses of $10^6 M_{\odot}$ can heat the disc. \cite{lace1984} extended the work to three dimensions and found that GMCs are quite efficient at redirecting the velocity of stars out of the plane, but rather inefficient at increasing its amplitude.

As a secular heating source, the Galactic bar has also been discussed in the literature \citep*{saha2010, gran2016}. However, although heating induced by the bar mainly takes place in the inner disc, the effect in the outer disc is not clear. Thus, the bar may contribute little to the observed AVR in the solar neighbourhood \citep{moet2016}.

Radial migration is another important mechanism in disc evolution \citep[e.g.][]{sell2002, rosk2008, scho2009b, minc2010, kord2015}, and has been proposed as a source of disc heating \citep[e.g.][]{scho2009a, loeb2011, rosk2013}. As \cite{sell2002} and \cite*{solw2012} pointed out, churning by spiral structures can redistribute the stars radially and hence causes far more mixing with very limited heating. Furthermore, \cite{minc2012} showed that the contribution of radial migration to the disc heating is negligible because of the balance between the outwards migrating stars, which `heat' the disc, and inward migrating stars, which `cool' it.

The shape of the velocity ellipsoid can also be associated with the heating processes providing that most of the stars are in equilibrium. \cite{lace1984} concluded that GMC scattering should cause that the value of vertical dispersion (\sigZ) is in between the radial (\sigR) and azimuthal (\sigPhi) components. However, this is inconsistent with the observed flattened velocity ellipsoid, i.e.,~$\sigma_{\rm z} < \sigma_{\phi}$ \citep{holm2009}. \cite{sell2008}, using simulations, clarified that Lacey's prediction neglected the perturbations from the distant encounters. \cite*{ida1993} corrected Lacey's heating rate and predicted that the vertical dispersion is the smallest among the three dispersion components. \cite{tian2015} showed that, for the stars located close to the Galactic mid-plane, vertical dispersion is the smallest component. \cite*{smit2012} found that although the shape of the velocity ellipsoid for the metal-rich stars is as expected, the ellipsoid for metal-poor stars is rounder, i.e.,~$\sigma_{\rm z} > \sigma_{\phi}$.

\cite{bovy2012} developed a ``mono-abundance population'' method (MAP) by analysing spatial structure for stars in small bins in \feh\ and \afe\ space. They found that the thick disc has larger scale height and smaller scale length than the thin disc. The distinct properties of the thick disc population were later confirmed by \cite{bovy2016, mack2017}, using a similar method. Although these results well presented how chemical structural parameters of discs vary with abundance and age, the lack of a direct connection between age and velocity ellipsoids leads to an indirect way of exploring the dynamical evolution of the disc structures.

In this paper, we use sub-giant branch (SGB) and low red giant branch (RGB) stars to revisit the AVR in the solar neighbourhood. With the measured radial velocities and ages from The Large Sky Area Multi-Object Fiber Spectroscopic Telescope (LAMOST) survey \citep{cui2012} DR3 catalogue, combined with the parallaxes and proper motions from the Tycho-Gaia Astrometric Solution \citep[TGAS;][]{gaia2016}, we are able to construct the age--kinematics relation within 1\,kpc of the Sun, which for this purpose is the solar neighbourhood.

The paper is organized as follows. In Section~\ref{sec:sample}, we describe our data and the methods to derive the age and the velocity ellipsoid. Then we show the main result in Section~\ref{sec:result}, and discuss the result in Section~\ref{sec:discussion}. Our final conclusion is given in Section~\ref{sec:conclusion}.

\section{The Samples}
\label{sec:sample}
\subsection{The LAMOST SGB/RGB stars}\label{sec:sgb}
LAMOST survey has well observed a few $10^5$ K giant stars~\citep{liu2014}. Recent studies showed that they can be used as age tracers~\citep{martig2016}. For LAMOST K giant stars, \cite{ho2017b} have provided ages for about 230,000 of them. However, because we want to use the accurate proper motions provided by TGAS, most of these K giant stars are located beyond 1\,kpc. Consequently, they are either not included in the TGAS catalogue or do not have reliable astrometric measurements.

Therefore, following \cite{liu2015}, we use the SGB/RGB stars as the age tracers. Specifically, stars with $5000<T_{\rm eff}<5300$\,K and $3<\log g<4$\,dex are selected for age estimation. The SGB stars and the low RGB stars (located at the base of the red giant branch) can well separate the age older than 1\,Gyr in $T_{\rm eff}$--$\log g$ plane. The SGB stars with ages younger than 1\,Gyr are difficult to isolate because they are located at larger $T_{\rm eff}$ and smaller $\log g$ and hence may be contaminated by horizontal-branch stars. Compared to the turn-off stars, another group of age tracers, the selected SGB/RGB samples are less affected by the main-sequence stars, fast-rotating stars, blue stragglers, and binaries.

\subsection{The age estimation}
We adopt the method to estimate the age for the SGB/RGB stars provided by \cite{liu2015}. We compare the LAMOST pipeline \citep{wu2011,wu2014} provided stellar parameters of these stars, $T_{\rm eff}$, $\log g$, and \feh, with PARSEC isochrones \citep{bressan2012}. The isochrones have a denser grid in logarithmic age ($\Delta \log a=0.01$) and metallicity ($\Delta {\rm [Fe/H]=0.1}$) with the coverage of $4950 \leq T_{\rm eff} \leq 5350$ and $2.9 \leq \log g \leq 4.1$.

For the $i$th star, the likelihood that it has stellar parameter set $O_{i}$=($T_{{\rm eff},i}$, $\log g_{i}$, ${\rm [Fe/H]}_{i}$) given age $\tau$, initial mass $\mathcal{M}_{\rm ini}$, and absolute magnitude $M$ can be written as
\begin{equation}\label{eq:likelihoodage}
\begin{aligned}
L_{i}(O_{i}|\tau,\mathcal{M}_{\rm ini},M)=&\\
&\exp\left(-\sum_{k=1}^{3}{(O_{i,k}-T_{k}(\tau,\mathcal{M}_{\rm ini}, M))^2\over{2\sigma_{i,k}^2}}\right),
\end{aligned}
\end{equation}
where $T_{k}(\tau,\mathcal{M}_{\rm ini}, M))$ is the predicted stellar parameter set ($k=1$, 2, and 3 correspond to $T_{{\rm eff},i}$, $\log g_{i}$, or ${\rm [Fe/H]}_{i}$, respectively) from isochrones and $\sigma_{i,k}$ the uncertainty of the observed stellar parameters.
As the nuisance parameters, $\mathcal{M}_{\rm ini}$ and $M$ are marginalized and leave $\tau$ as the only unknown parameter.

\cite{liu2015} assessed the performance of the age estimates with the test mock data and concluded that 1) the typical random error of the age is about 30\% and 2) it may be underestimated by at most 2\,Gyr when the age estimate is larger than 8\,Gyr. Therefore, we adopt 30\% uncertainty as the age estimate for all the sample stars.

\subsection{The Sample Selection}
61,134 SGB/RGB stars following the selection criteria mentioned in section~\ref{sec:sgb} are selected from LAMOST DR3, which in total contains more than 4 million stellar spectra with stellar parameters been estimated. The accuracy of the radial velocity of LAMOST is better than 5\,\kms\ \citep{gao2014} with a systematic offset of 5.7\,\kms\ according to \cite{tian2015}.

As part of GAIA survey DR1 \citep{gaia2016}, the TGAS catalogue provides parallaxes and proper motions for about 2 million stars down to the 12th mag. The uncertainty of the proper motions is about 1\,${\rm mas\,yr^{-1}}$, while the typical uncertainty for the parallaxes is 0.3\,mas including systematic errors.

We cross-match the LAMOST SGB/RGB stars with TGAS catalogue and obtain a sample of 3,564 common stars with age estimates. Following~\citet{liu2017}, the LAMOST targets are selected purely based on their photometry. The TGAS selection effect correction is mostly due to uneven spatial distribution coverage in the 14 month \emph{Gaia} data release. Therefore, both surveys have a negligible selection effects in the kinematics.

Because the distance directly converted from the inverse of the parallax may be significantly overestimated for stars farther than 500\,pc, we adopt the distance derived from \cite{astr2016}, who applied a Bayesian method to the distance estimation from the parallax with the Milky Way prior. Note that the systematic bias of the TGAS astrometry has been applied twice unintentionally by the authors (Bailer-Jones private communication). This may induce about 1.5\,\kms\ in the uncertainty of the velocity. However, such a small value may not substantially change the velocity ellipsoids since the measurement uncertainty is contributed as a squared term in the latter quantities.

Figure~\ref{fig:zhist} shows the vertical spatial distribution of the sample, which roughly spans to $|z| \sim 1000$\,pc, allowing for the comparison of the kinematics at different vertical heights.

\subsection{Likelihood function}
We adopt cylindrical coordinates to decompose the stellar motion into radial, azimuthal and vertical velocity components. The position of the Sun is adopted as $(X, Y, Z) = (8000, 0, 27)$\,pc, while the solar motion with respect to the local standard of rest (LSR) is adopted as $(U_{\odot}, V_{\odot}, W_{\odot}) = (9.58, 10.52, 7.01)$\,\kms \citep{tian2015}.

We assume that the velocity distribution in the disc can be modelled as a multivariate Gaussian distribution, but note that the distribution of the azimuthal component need not be a Gaussian. \cite{tian2015} found that the Gaussian approximation does not induce significant systematic bias in the estimation of the velocity ellipsoid. Thus, the likelihood function can be constructed such as
\begin{equation}
	L = \prod_{i=1}^{N} { \frac {1} {(2\pi)^{\frac {n} {2}} \left|{\bm{\Sigma}}\right|^{\frac {1} {2}}} \exp \left ( - \frac {1} {2} \bm{(v - \mu)}^{T} \bm{\Sigma}^{-1} \bm{(v - \mu)} \right ) },
\label{eqn:likelihood}
\end{equation}
where $\bm{\Sigma}$ is the squared sum of the intrinsic velocity tensor and the covariance matrix of the uncertainty of the velocity estimates and $\bm{\mu}$ represents for $(v_{\rm R}, v_{\rm \phi}, v_{\rm z})^{T}$ in the Galactocentric cylindrical coordinates. $\bm{\mu}$ and the terms related to the intrinsic velocity ellipsoid in $\bm{\Sigma}$ are the parameters to be determined. We choose the prior of the velocity dispersions to be uniformly distributed in the range of $(0, +\infty)$. Note that the uncertainties of the observed line-of-sight velocity ($\approx 5$\,\kms), the proper motions, and the Bayesian distances have been taken into account when computing $\bm{\Sigma}$. Eq.~\ref{eqn:likelihood} is equivalent to the case that the measurement uncertainties are deconvolved from the measured dispersions so that the intrinsic velocity ellipsoid would not be broadened by the measurement errors.

We estimate the mean velocities and the velocity ellipsoids for a given stellar population using the Markov chain Monte Carlo (MCMC) simulation with the \emph{emcee} package \citep{fore2013}.

\section{Results}
\label{sec:result}

\subsection{Separations in \z\ and age}

\begin{figure}
	\centering \includegraphics[scale=0.55]{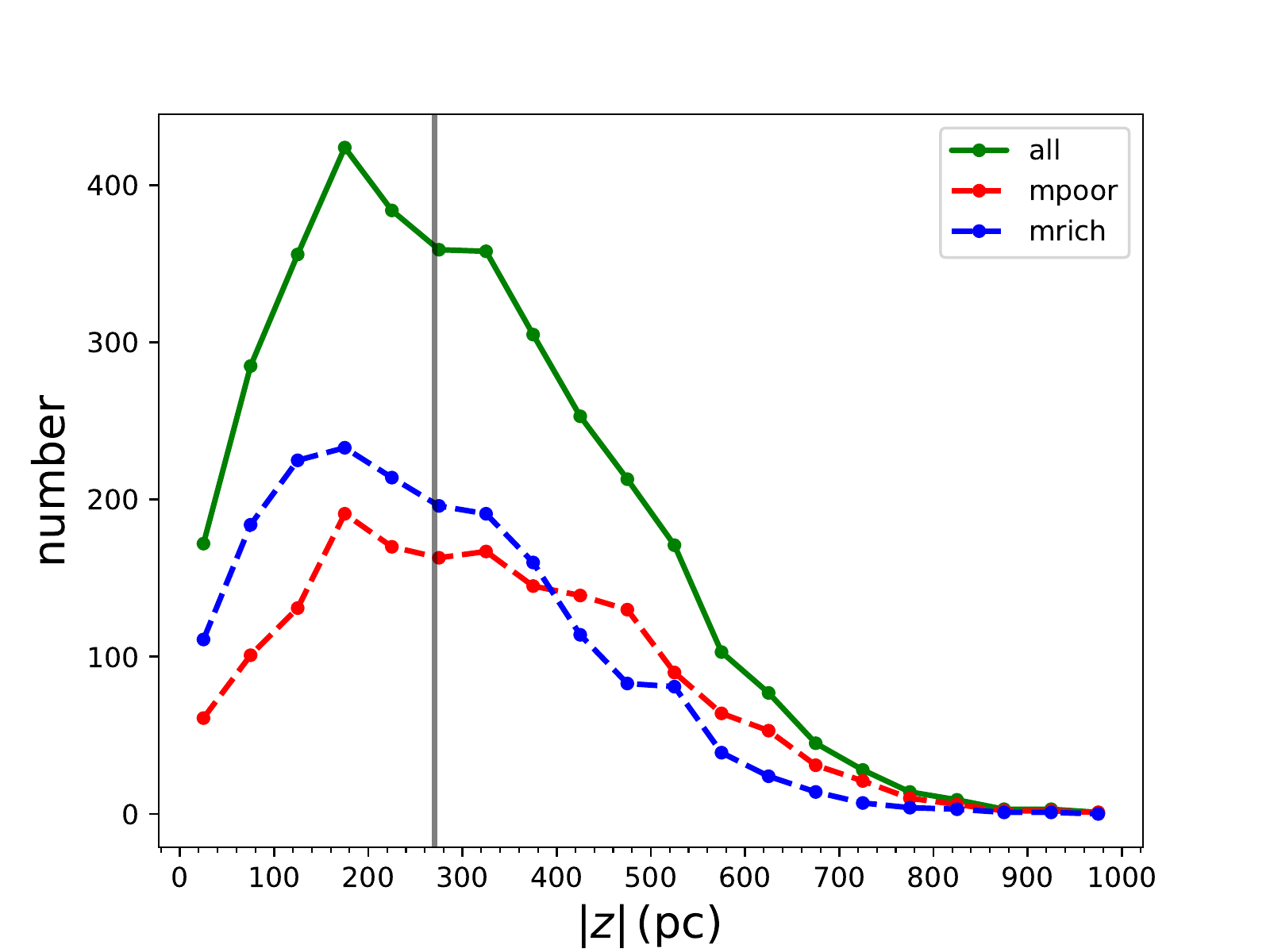}
	\caption{The vertical spatial distribution of the SGB/RGB star samples. The green line represents for the distribution for the whole sample, while the red and blue lines stand for the distributions for the sample with ${\rm [Fe/H]<-0.2}$\,dex and ${\rm [Fe/H]>-0.2}$\,dex, respectively. The black vertical line indicates the point of $|z|=270$\,pc to separate the data into two roughly equal number sub-groups (see the text for details).}
	\label{fig:zhist}
\end{figure}

The distribution of the sample in the vertical height \z\ is displayed in Figure~\ref{fig:zhist}. It has a broad distribution with a peak located at around 100-200\,pc. The largest vertical distance for the sample reaches to about 1000\,pc. This range of \z\ allows for the detection of both the thin and thick discs. In order to compare the AVR at different heights, the samples are divided into two different \z\ bins such that each \z\ bin contains similar number of stars. The sample with $|z|<270$\,pc has 1,775 stars, while the sample with $|z|>270$\,pc has 1,789 stars.

The stars located in each \z\ bin are split into various age bins. Table~\ref{age_bin1} and~\ref{age_bin2} list the age bins for the whole dataset, the metal-rich sample, and the metal-poor sample (the definition of the two metallicity separated samples can be found in section~\ref{sec:metal}) at $|z|<270$\,pc and $|z|>270$\,pc, respectively. The principle to split the data into age bins is that the width of any age bin should be 30\% around its central value according to the uncertainty of the age estimates \citep{liu2015}. Meanwhile, the centre points of the neighbouring age bins should be separated by at least 15\% of the age value. Furthermore, the number of stars in each age bin is set to be between 100 and 400. Consequently, the neighbouring age bins may overlap with each other, which smooth the resulting velocity ellipsoids such that the arbitrarily fluctuation in the AVR can be reduced.

\begin{table*}
	\caption{Kinematical features for stars with $|z| < 270$\,pc at various age bins.}
	\label{age_bin1}
	\begin{center}
		\input{tab1.tex}
	\end{center}
\end{table*}

\begin{table*}
	\caption{Kinematical features for stars with $|z| > 270$\,pc at various age bins.}
	\label{age_bin2}
	\begin{center}
		\input{tab2.tex}
	\end{center}
\end{table*}

\subsection{Separations in metallicity}
\label{sec:metal}
The stellar populations can be well characterized by the stellar age, metallicity, and $\alpha$-abundance. The kinematical features for the populations with various ages, metallicities, and $\alpha$-abundance may hint at the dynamical evolution of the discs. However, for the data in this work, the number of stars in each age bin is set to be between 100 and 400 to meet the requirement of statistical significance of the kinematics with enough age resolution. Within each age bin, the number of stars may not be sufficient for further dividing into many metallicity bins. Moreover, not all of the sample stars, but only about two thirds have \afe\ been estimated \citep{ho2017}. Therefore, we decide to only separate the data into two sub-groups in \feh\ so that the variation of the AVR can be still investigated for the chemically-defined thin/thick disc. The chemical definition of the thin and thick discs is usually based on \afe. The stars with larger \afe\ are likely from the thick disc population, while those with smaller \afe\ are likely from the thin disc. The separation point is usually selected at around ${\rm [\alpha/Fe] \sim 0.2}$\,dex \citep{lee2011}.

\begin{figure*}
	\centering \includegraphics[width=1.0\linewidth,angle=0]{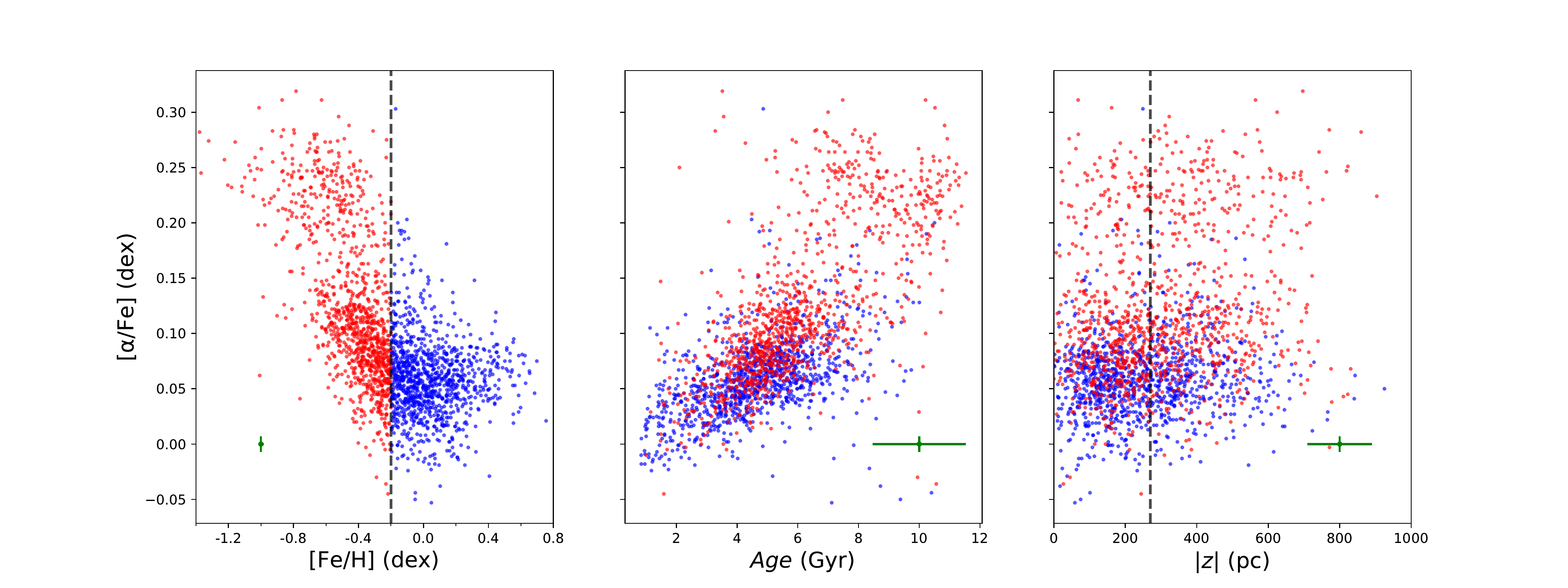}
	\caption{\afe-\feh\ relation (left panel), \afe-age relation (middle panel) and \afe-$|z|$ relation (right panel) for part of (2,139 stars) our sample (3,564 stars). In the left panel, the vertical black dashed line at ${\rm [Fe/H] = -0.2}$\,dex gives the dividing criterion for the metal-poor and metal-rich populations. The metal-poor population is dominated by $\alpha$-enhanced star, while most of stars in the metal-rich population have lower $\alpha$-abundance. The middle panel shows that the $\alpha$-enhanced stars are mostly old, while $\alpha$-poor stars spread in a wide range of age. The right panel shows that both metal-poor and metal-rich populations distribute broadly in vertical direction, at both sides of $|z| = 270$\,pc, which is indicated by the vertical dashed line. The green error bars in each panel give the corresponding median uncertainties of the sample. Note that the median uncertainty of \feh\ is 0.02\,dex, which is almost invisible in the plot.}
	\label{fig:afeh}
\end{figure*}

Figure~\ref{fig:afeh} shows the age--\feh--\afe--\z\ relationship for a subset of 2,139 stars, which contain \feh\ and \afe\ measurements by \cite{ho2017}, from the whole 3,564 SGB/RGB star sample. \cite{ho2017} estimated the chemical abundances by cross-calibrating the LAMOST DR2 K giant stars with APOGEE data by applying \emph{The Cannon} technique. As expected, it shows a clear bimodality with a gap at around ${\rm [\alpha/Fe] \sim 0.15}$ - $0.2$\,dex. Following the criterion for \afe\ from previous works \citep{lee2011, liu2012}, we could separate the sample into $\alpha$-enhanced and -poor populations to represent the chemically-defined thick and thin discs. However, because \afe\ is not estimated for about one third of our samples, we have looked for an alternative criterion for the thin/thick disc separation to take the place of \afe.

\begin{figure}
	\centering \includegraphics[width=1.0\linewidth,angle=0]{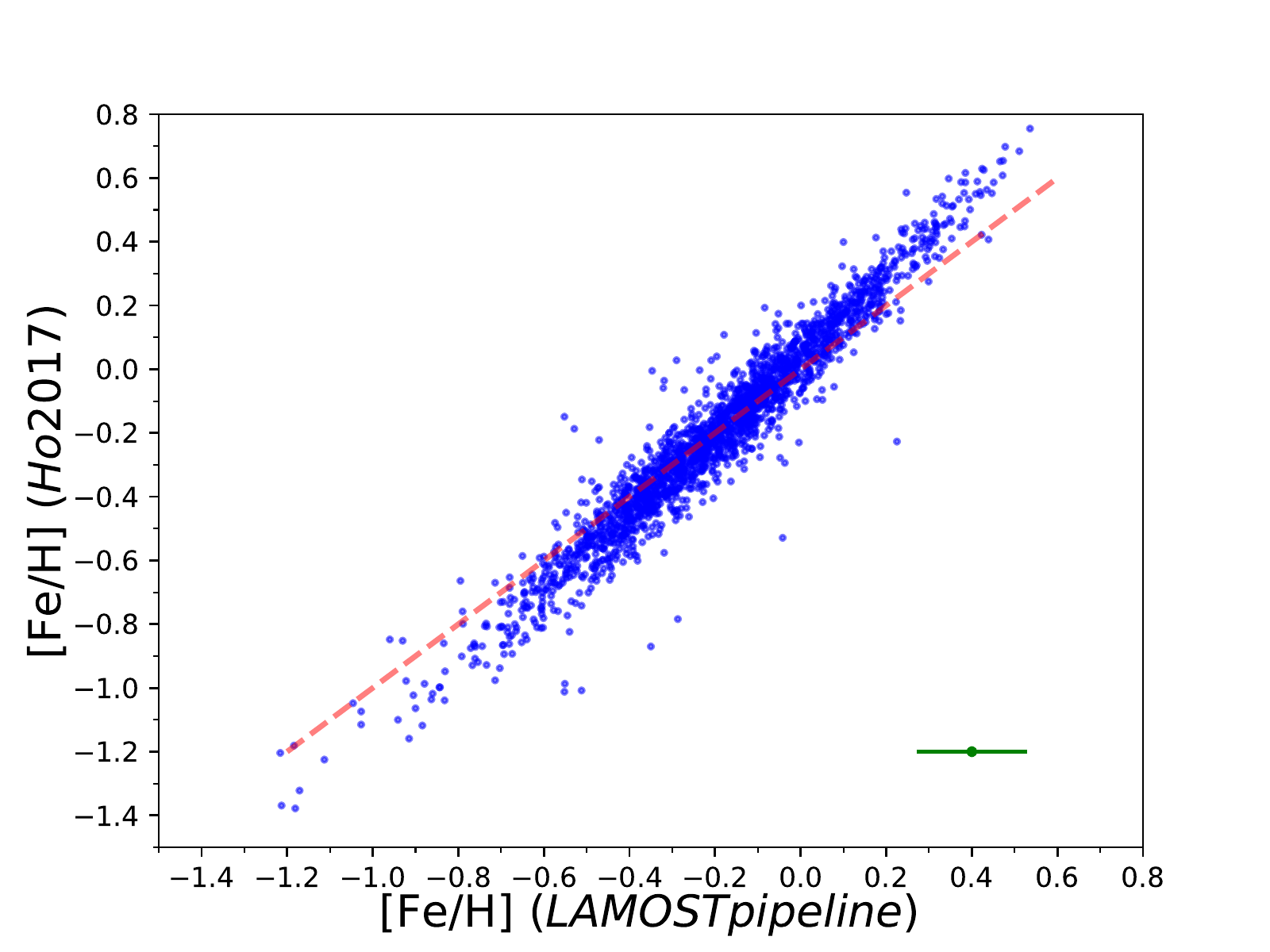}
	\caption{The \feh\ derived by \citep{ho2017} (labeled as Ho2017) versus the \feh\ from LAMOST pipeline (labeled as LAMOST pipeline). The red dashed line shows the $1:1$ relation. The green errorbar gives the median uncertainties.}
	\label{fig:feh2_feh3}
\end{figure}

We find that the two discs can be at least partly disentangled using ${\rm [Fe/H]=-0.2}$\,dex. On one hand, stars with ${\rm [Fe/H]>-0.2}$\,dex are only composed of those with ${\rm [\alpha/Fe] \lesssim 0.2}$\,dex, i.e.~only the chemically-defined thin disc contributes to this sub-group. On the other hand, the sample with ${\rm [Fe/H]<-0.2}$\,dex contains both $\alpha$-poor and $\alpha$-enhanced stars, implying that this sub-group may reflect the kinematical features for both thin and thick discs.

It is noted that, as shown in the middle panel of the figure, the stars with ${\rm [\alpha/Fe] \gtrsim 0.2}$\,dex from the metal-``poor" (${\rm [Fe/H]<-0.2}$) sub-group (red dots) dominate the regime of age $>7$\,Gyr, while those with ${\rm [\alpha/Fe] \lesssim 0.2}$\,dex are mostly 4 to 8\,Gyr old. Hence, the metal-poor sub-sample can be separated into two parts: 1) those with age $\gtrsim7$\,Gyr are dominated by the thick disc population; 2) those with age $\lesssim7$\,Gyr should be thin disc stars mostly older than 3\,Gyr. Meanwhile, the metal-``rich" (${\rm [Fe/H]>-0.2}$) sub-group (blue dots) is prominent in the range of age $<8$\,Gyr with few stars extending to older age. This means that the chemically-defined thin disc can be traced by the metal-rich stars in the whole range of age. Therefore, although the separation at ${\rm [Fe/H]=-0.2}$\,dex is not perfect for disentangling the thin and thick discs, the significant difference in the distribution in the age--\afe\ plane for the two sub-groups is sufficient in investigating the different AVRs for the thin and thick discs. In the rest of the paper, we treat the metal-poor sub-sample with age $\gtrsim7$\,Gyr as the tracer of the chemically-defined thick disc and treat the metal-rich sub-sample as the probe of the thin disc population.

Finally, the right panel of the figure shows that both the metal-poor and metal-rich sub-sample are broadly distributed in both $|z|<270$ and $>270$\,pc. Hence the AVR at different vertical heights for both sub-sample can be compared.

In the rest of this paper, we use \feh\ derived from LAMOST data pipeline. As shown in Figure~\ref{fig:feh2_feh3}, the difference of \feh\ between \cite{ho2017} and the LAMOST pipeline is mostly within $0.1$\,dex. Therefore, we can still use \feh$=-0.2$\,dex to separate the samples into metal-poor and -rich sub-groups.

\subsection{The resulting AVRs}
\begin{figure*}
	\centering \includegraphics[width=1.0\linewidth,angle=0]{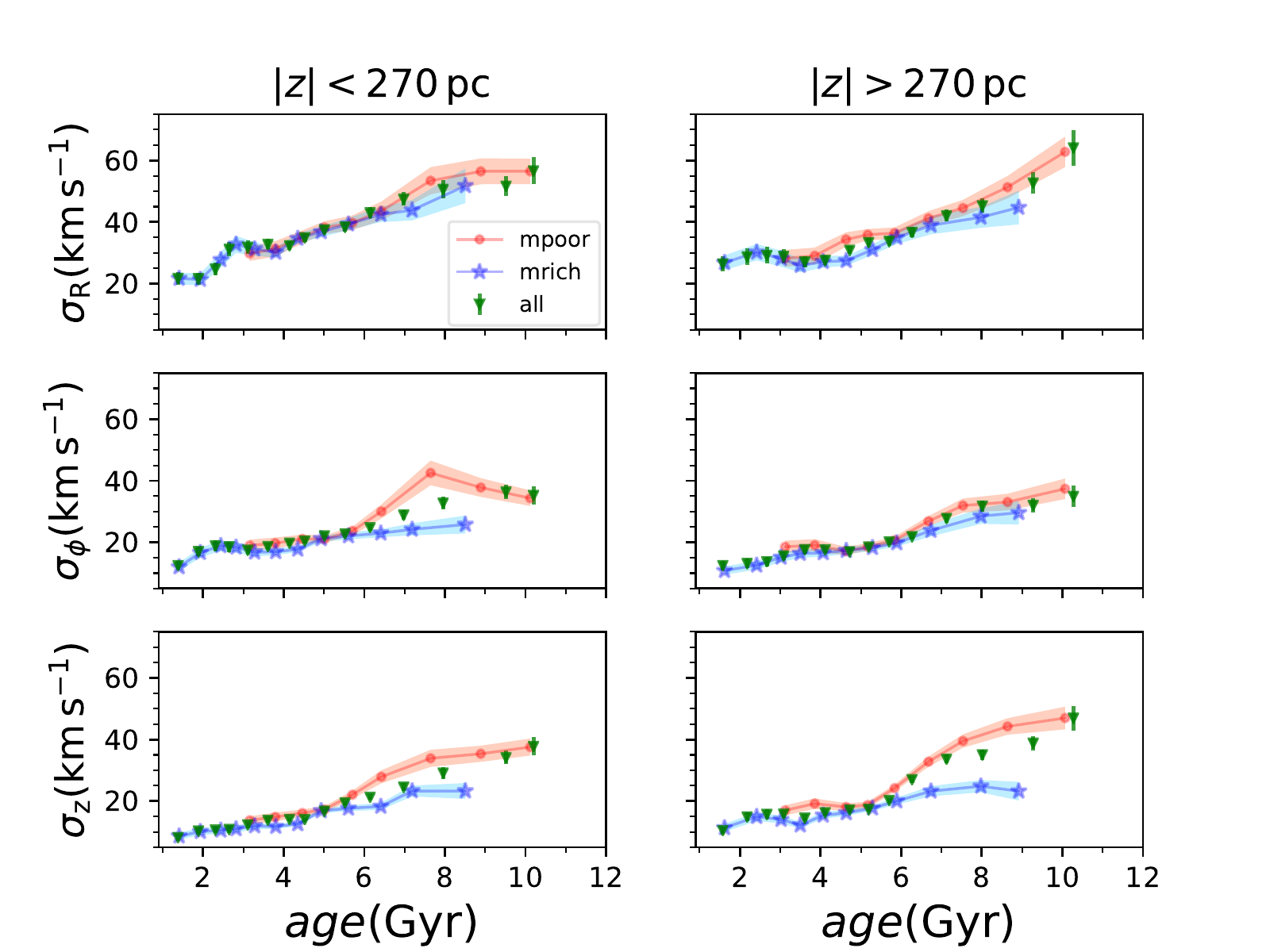}
	\caption{The AVR for the two sub-samples with different metallicities at two different $|z|$ bins. Different coloured symbols represent different samples. The green triangles represent the whole sample, labelled as ``all''. The red dots represent the metal-poor sample, labelled as ``mpoor''. The blue stars represent the metal-rich sample, labelled as ``mrich''. The red and blue shaded bands give the corresponding errors for the metal-poor and metal-rich sample, respectively.}
	\label{fig:avr}
\end{figure*}

For each \z\ and \feh\ bin, the 3-dimensional velocity ellipsoids for the populations with various ages are determined with the MCMC simulation. The diagonal components of the velocity ellipsoids, i.e.~\sigR, \sigPhi, and \sigZ, as functions of age for each \z\ and \feh\ bin are displayed in different panels of Figure~\ref{fig:avr}. Meanwhile, the velocity dispersions, their ratios, the vertex deviations, and the tilt angles are listed in Table~\ref{age_bin1} and~\ref{age_bin2} for stars located at $|z|<270$ and $>270$\,pc, respectively.

The left hand panels of Figure~\ref{fig:avr} show the trends of \sigR, \sigPhi, and \sigZ\ with age from top to bottom, respectively, for $|z|<270$\,pc. Looking at the velocity dispersions for the two sub-samples (red dots for metal-poor stars and blue stars for metal-rich), all of the three dispersions increase with age. This is consistent with previous works \citep[e.g.][]{bour2009, holm2009, smit2012, tian2015}. However, the increasing rates with age for the two sub-samples are significantly different. Firstly, the red dots, which represent the metal-poor sub-group stars, show plateaus of velocity dispersions substantially larger than those for the metal-rich stars for age $\gtrsim7$\,Gyr. They are more significant in \sigPhi\ and \sigZ\ than in \sigR. For the metal-poor stars, \sigPhi\ is between 34 and 43\,\kms\ for age $\gtrsim7$\,Gyr, while it is only around $24\sim26$\,\kms\ for the metal-rich stars at the same ages. Similarly, for the metal-poor stars, \sigZ\ is around $34\sim38$\,\kms\ for age $\gtrsim7$\,Gyr, while for the metal-rich stars, the value is only at $18\sim23$\,\kms. Since the metal-poor sub-sample with age $\gtrsim7$\,Gyr is dominated by the chemically-defined thick disc, it seems that the significant differences are caused by the different kinematical features between the thick and thin disc. Secondly, for age $<7$\,Gyr, velocity dispersions for the metal-poor sub-sample and those for metal-rich sub-sample are quite similar. These trends can also be seen from Table~\ref{age_bin1}. Finally, at around 2-3\,Gyr, both \sigR\ and \sigPhi\ show an abrupt jump-up, which is not obviously seen in \sigZ\ at the same age.

The right hand panels of Figure~\ref{fig:avr} display the AVRs for the three velocity dispersions for $|z|>270$\,pc. Almost all the differences between the metal-poor and metal-rich sub-groups shown in the lower vertical height in the left panels are found in the sample for $|z|>270$\,kpc. For the metal-rich sub-sample, the trends of \sigR, \sigPhi\ and \sigZ\ are quite similar with those for $|z|<270$\,pc. In other words, \sigR, \sigPhi\ and \sigZ\ do not show significant vertical gradients in the metal-rich sub-sample. The values of \sigR\ and \sigPhi\ for the metal-poor sub-sample are also quite similar on both side of $|z|=270$\,pc. However, the values of \sigZ\ for age $\gtrsim7$\,Gyr and $|z|>270$\,pc are significantly larger than those with similar ages but $|z|<270$\,pc for the metal-poor sample.

\section{Discussion}
\label{sec:discussion}

\subsection{Thin and thick discs}

As shown in Figure~\ref{fig:avr}, the significant difference in velocity dispersions between the metal-rich and metal-poor sub-samples for age $\gtrsim7$\,Gyr indicates that the chemically-defined thick disc (traced by the meta-poor sub-samples with age $\gtrsim7$\,Gyr) has distinct kinematic features, i.e.~it is hotter than the old thin disc stars (traced by the metal-rich sub-samples) in all three velocity dispersions, especially in vertical direction. This implies that although the heating of the thin disc leads to a relatively larger \sigZ, i.e.~makes it thicker, at early time, it can not produce such a large \sigZ\ plateau as the chemically-defined thick disc. Therefore, it seems that the chemically-defined thick disc should be a distinct population to the heated old thin disc. Note that, for age $\gtrsim7$\,Gyr, our age estimates may suffer from larger systematics as well as larger uncertainties \citep{liu2015}, hence the lower age limit for the thick disc could be slightly larger than 7\,Gyr.

Moreover, it is noted that although the thick disc traced by the metal-poor sub-sample with age $\gtrsim7$\,Gyr shows large difference in \sigZ\ with the thin disc, its \sigR\ and \sigPhi\ are only moderately larger than the thin disc with similar age.

Comparing \sigPhi\ for $|z|<270$ with that at $|z|>270$\,pc, \sigPhi\ does not change very much at different heights for the thick disc stars, while it increases from $\sim25$\,\kms\ to about $\sim30$\,\kms\ for the thin disc sub-sample with age $\gtrsim7$\,Gyr. This implies that the thick disc should be formed on relatively smaller time scale such that the whole population shows similar velocity dispersion, while the thin disc may be formed on a longer time scale and thus the secular dynamical evolution can play some role to heat and scatter the older stars to a larger range of height.

Finally, for the metal-poor sub-sample with age $<7$\,Gyr, because the chemically-defined thin disc dominates in these data, the stars essentially show similar AVRs to the metal-rich sub-sample in all three components.

Although the significantly different velocity dispersions shown in AVR can separate the thick from the thin disc, this remains insufficient for discriminating the origin of the thick disc. Using cosmological simulations, \cite*{mart2014} found that galaxies commonly undergo an active phase of mergers at high redshift, creating a thick stellar component. In this scenario, the oldest stars with age $> 9$\,Gyr are born kinematically hot.

\subsection{Heating in the disc}

At lower redshift, \cite{mart2014} showed that the vertical velocity dispersion (\sigZ) increases smoothly with age. They suggested that the disc heating is mainly contributed by disc growth and a combination of spiral arms and bars coupled with overdensities in the disc and vertical bending waves. The heating mechanisms lead to a power-law like AVR, while the active phase of mergers at high redshift breaks the power-law like AVR with age $> 9$\,Gyr. The authors pointed out that the jumps in AVR may also be created by mergers \cite[also see][]{hous2011}. However, they found that errors of 30\% in stellar ages can significantly blur the signatures of mergers. Therefore, we apply the power-law relation to the chemically-defined thin disc component.

\begin{figure*}
	\centering \includegraphics[width=0.9\linewidth,angle=0]{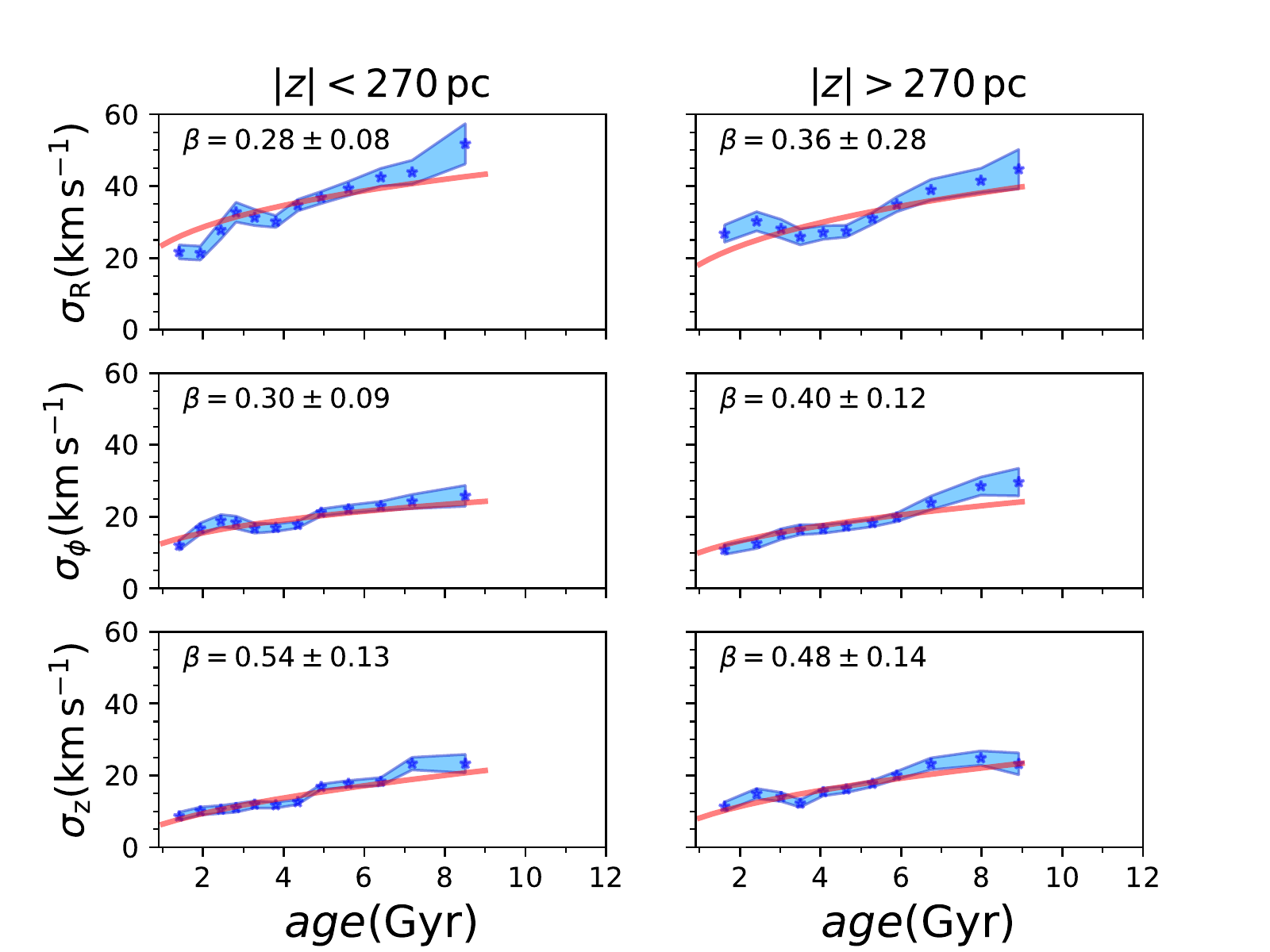}
	\caption{AVRs at different vertical heights for metal-rich population (${\rm [Fe/H] > -0.2}$\,dex). Blue stars show the derived velocity dispersion in different age-bins, while the blue shaded bands indicates the corresponding errors. The red solid curve shows the best fit power law for bins with age between 3\,Gyr and 9\,Gyr.}
	\label{fig:fit}
\end{figure*}
We fit the AVR for the metal-rich stars within 3 Gyr - 9 Gyr using a power-law relation. The samples not in the range of age suffers from larger uncertainties due to the smaller numbers in these age bins. The best-fit models of the AVR for the metal-rich populations are shown in Figure~\ref{fig:fit}. The derived AVRs in all three directions are well represented by a power-law when the youngest and oldest bins are excluded.

The derived power indices for the radial dispersions for $|z|<270$ and $|z|>270$\,pc are $\beta = 0.34 \pm 0.17$ and $\beta = 0.39 \pm 0.13$, respectively, while for the azimuthal direction we obtain $\beta = 0.34 \pm 0.12$ and $\beta = 0.42 \pm 0.14$. The derived power indices for the vertical direction, which are $\beta = 0.56 \pm 0.14$ and $\beta = 0.51 \pm 0.15$, are higher than the other two directions. The radial and azimuthal velocity dispersion are approximately coupled by the relation:
\begin{equation}
	\frac {\sigma_{\rm R}} {\sigma_{\phi}} = \frac {\kappa} {2\Omega},
	\label{eqn:rphi}
\end{equation}
where $\kappa$ and $\Omega$ are the epicycle frequency and the azimuthal frequency for nearly circular orbits in the solar neighbourhood \citep{bt2008}. Thus, the in-plane power indices are expected to be similar. Furthermore, the in-plane power indices at high-\z\ are slightly higher than that in low-\z, indicating that the heating sources might be more scattered to higher-\z. Nevertheless, all these in-plane power indices are around 0.3-0.4, while the vertical power indices are around 0.5, which is consistent with other works \citep[e.g.][]{aume2009, holm2009}.

We conclude that the AVR of the metal-rich population is consistent with heating mechanisms via scattering processes mainly induced by spiral arms and GMCs.

Note that the resulting $\bm{\sigma}$ for the thin disc are systematically higher than the fitted power-law at age $>7$\,Gyr. This could be because that the older populations are more affected by larger uncertainties from the age estimates. According to \cite{liu2015}, the ages can be underestimated by at most 2\,Gyr in the old regime. Then the old stars with larger velocity dispersions may be merged to the bins of slightly younger age. As a result, the slope of the age-$\sigma$ relation at the old age regime becomes steeper. An alternative reason may be that the heating process is not constant on long time scales. Considering that the number density of GMCs was significantly higher in the past than in the present day \citep*{chia1997}, older stars may therefore be more efficiently scattered in the early time. It is also possible that the more frequent mergers at high redshift can affect the velocity dispersion.

\subsection{$\sigma$-ratio}
\begin{figure*}
	\centering \includegraphics[width=0.9\linewidth,angle=0]{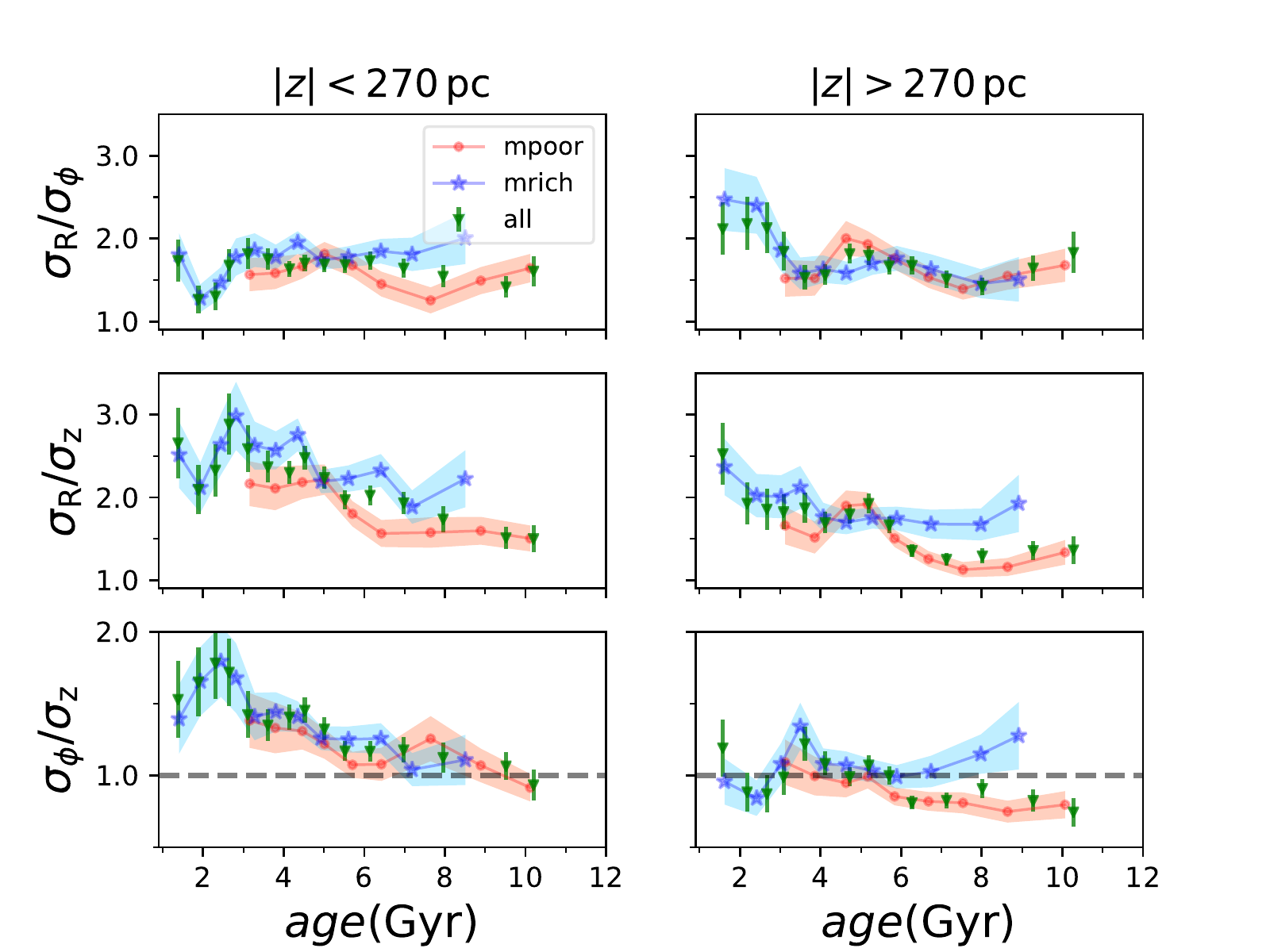}
	\caption{Ratios of velocity dispersion at different vertical heights. Green triangles show the ratio for the whole sample. The red dots and blue stars with shaded errors correspond to stars with ${\rm [Fe/H] < -0.2}$\,dex (metal-poor) and ${\rm [Fe/H] > -0.2}$\,dex (metal-rich), respectively.}
	\label{fig:ratio}
\end{figure*}

The velocity dispersion ratios of our sample are shown in Figure~\ref{fig:ratio}. There is no substantial trend in the ratio $\sigma_{\rm R}/\sigma_{\phi}$ with age. The result is consistent with the theoretical expectation that the ratio $\sigma_{\rm R}/\sigma_{\phi}$ is a constant since they are coupled (see Eq.~\ref{eqn:rphi}). On the other hand, because the stars are more efficiently heated in vertical direction, as can be inferred from the larger power indices in vertical direction, the ratios of in-plane to vertical dispersions decrease with age.

Previous observations show that the vertical dispersion \sigZ\ is smallest among the three components. \cite{lace1984} analysed the scattering by GMCs and predicted that the value of \sigZ\ should be in between \sigR\ and \sigPhi. \cite{ida1993} later found that \cite{lace1984} overestimated the vertical heating rate while underestimated the radial heating rate. They predicted that the vertical component should be the smallest although precise shape depends on the local slope of the rotation curve. Their expectations are confirmed by simulations by \cite{shii1999}. \cite{sell2008}, using simulations, pointed out that the ratio $\sigma_{\phi}/\sigma_{\rm z}$ depends on the distant GMCs, that is, when enough distant GMC perturbations are taken into account, \sigZ\ generally remains the smallest of the velocity dispersions.

However, we find that the ratio $\sigma_{\phi}/\sigma_{\rm z} \approx 1$ for stars with $|z| > 270$\,pc. Moreover, it drops below 1 for metal-poor stars with $|z| > 270$\,pc. In other words, the vertical dispersion for the metal-poor stars at higher-\z\ is larger than the azimuthal dispersion. \cite{holm2009} showed a flattened shape of velocity ellipsoid with \sigZ\ to be the smallest, although the ratio $\sigma_{\rm z}/\sigma_{\phi}$ increases with age. \cite{smit2012} confirmed a similar velocity ellipsoid shape for the metal-rich stars (${\rm [Fe/H] \ge -0.5}$\,dex). However, they found that the value of \sigZ\ is in between the other two for the metal-poor stars (${\rm [Fe/H] < -0.5}$\,dex), which is very similar to our result. \cite{tian2015} also reported that vertical velocity dispersion values are in the middle for stars with $T_{\rm eff} \sim 6500$\,K at $300 < |z| < 500$\,pc, which are likely dominated by the old turn-off thick disc stars. Therefore, it is likely that the ratio $\sigma_{\phi}/\sigma_{\rm z} < 1$ is due to the distinct kinematic feature of the dominating old thick disc population.

\subsection{Vertex deviation and tilt angle}
The covariances (or the cross-terms) of the velocity ellipsoid reflects the tilts of the ellipsoid, which is associated with the non-axisymmetric structures in the Galactic gravitational potential. They are usually quantified by tilt angle, which is the deviation from the ${\rm R-z}$ plane, and vertex deviation, which is the deviation from the ${\rm R-\phi}$ plane, for convenience.
The definition of vertex deviation is:
\begin{equation}
	l_{\rm v} \equiv \frac {1} {2} \arctan \left ( \frac {2\sigma_{\rm {R} \phi}^{2}} {\sigma_{\rm R}^{2} - \sigma_{\phi}^{2}} \right ),
\end{equation}
and the definition of the tilt angle is:
\begin{equation}
	\alpha \equiv \frac {1} {2} \arctan \left ( \frac {2\sigma_{\rm R z}^{2}} {\sigma_{\rm R}^{2} - \sigma_{\rm z}^{2}} \right ).
\end{equation}

\begin{figure}
	\centering \includegraphics[width=1.0\linewidth,angle=0]{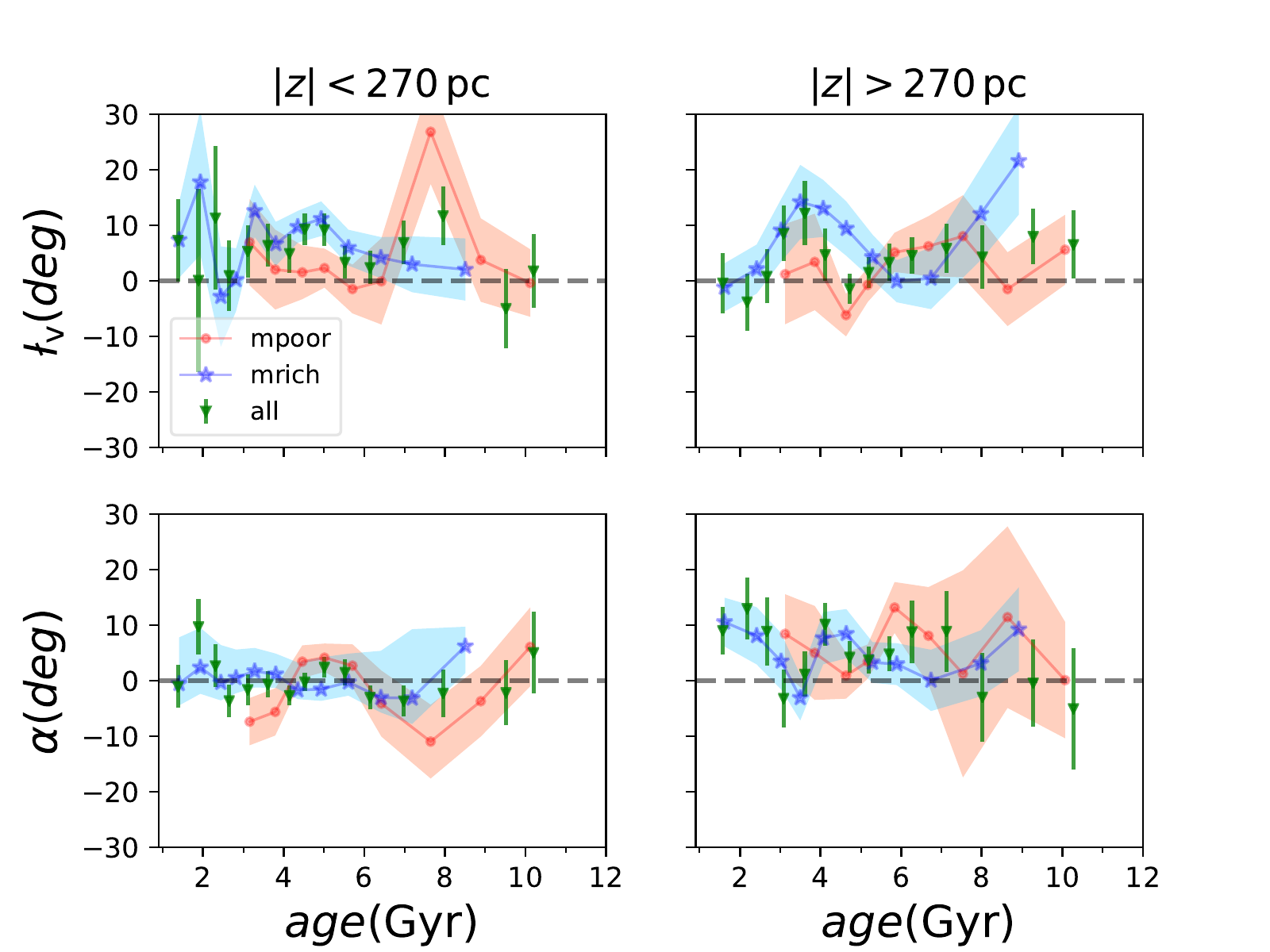}
	\caption{The vertex deviation and the tilt angle as functions of age at different vertical heights. Green triangles show the ratio for the whole sample. The red dots and blue stars with shaded errors correspond to stars with ${\rm [Fe/H] < -0.2}$\,dex (metal-poor) and ${\rm [Fe/H] > -0.2}$\,dex (metal-rich), respectively.}
	\label{fig:tilt}
\end{figure}

The derived vertex deviation and the tilt angle at various age bins are shown in Figure~\ref{fig:tilt}. The vertex deviations at different vertical heights are plausibly zero for all stars but with large uncertainties. \cite{dehn1998} showed the vertex deviation as a function of colour index. The stars with $B-V>0.4$ were found to have vertex deviation at around $10^{\circ}$ with a $5^{\circ}$ uncertainty, while the stars with $B-V<0.4$ have vertex deviation around $20^{\circ}$--$30^\circ$ with an error between $2.8^{\circ}$ and $5.3^{\circ}$. \cite{smit2012} reported similar results for the metal-rich stars (${\rm [Fe/H] \ge -0.5}$\,dex). They also found that the vertex deviation is around $0^{\circ}$ for metal-poor stars but with large uncertainty. \cite{tian2015} found that the vertex deviations are around $0^{\circ}$ with an error between $5^{\circ}$ and $10^{\circ}$ for different effective temperatures at different $|z|$, although they found it can be as large as $\sim 27^{\circ}$ for stars with $T_{\rm eff} < 4500$\,K at $300 < |z| < 500$\,pc, yet the uncertainty is as large as $\sim 14^{\circ}$.

The tilt angle of our sample is around $0^{\circ}$ for most of the stars but the measurement has very large uncertainties. \cite{tian2015} found the tilt angle to be around $0^{\circ}$ with an error of $\sim 5^{\circ}$, while \cite{smit2012} reported a non-zero $\alpha$ to be around $10^{\circ}$ with an uncertainty of $\sim 10^{\circ}$. \cite{sieb2008} used red clump giants from the Radial Velocity Experiment (RAVE) survey and gave a tilt angle of $7.3^{\circ} \pm 1.8^{\circ}$ at $z \sim 1\,{\rm kpc}$ below the plane. \cite*{bude2015} used G-dwarfs which spans a large vertical range, reaching as far as 3\,kpc, and found an increasing trend of tilt angle with $|z|$ in the range of $0^{\circ} - 15^{\circ}$ with a $10^{\circ}$ uncertainty.

\section{Conclusions}
\label{sec:conclusion}

In this paper, we use 3,564 SGB/RGB stars selected from LAMOST DR3 and TGAS data to derive the AVR in the solar neighbourhood. By dividing the samples into two sub-groups with ${\rm [Fe/H]<-0.2}$\,dex and ${\rm [Fe/H]>-0.2}$\,dex, we are able to compare the AVR between the thin and thick discs. We find that:
\begin{enumerate}
\item{the large difference in the AVR between the metal-poor and metal-rich sample at age $\gtrsim7$\,Gyr favours that the chemically-defined thick disc is a distinct population formed within a short time scale, which is different from the thin disc.}
\item{the AVR for the metal-rich stars follows a power-law with best-fit indices between 0.3 and 0.4 for in plane velocity dispersions and around 0.5 for vertical dispersion. This is consistent with secular heating of the thin disc which is mainly driven by spiral arms and GMCs. As \cite{casa2011} and \cite*{aume2016} point out, large age uncertainties can flatten the AVR, therefore the intrinsic increasing rate of velocity dispersion should be higher.}
\item{$\sigma_{\rm R} / \sigma_{\rm z} < 1$ for the old metal-poor stars, which belong to the thick disc. For the metal-rich stars, which are mostly from the thin disc, the in-plane to vertical dispersion ratios decrease with age but remain $>1$ for the oldest stars.}
\item{the vertex deviations and the tilt angles are plausibly around zero, but with large uncertainties.}
\end{enumerate}

\section*{Acknowledgments}
We thank Christopher Flynn for the detailed and constructive review, which improved the paper a lot.
This work is supported by the National Key Basic Research Program of China 2014CB845700. C.L. acknowledges the National Natural Science Foundation of China (NSFC) under grants 11373032 and 11333003.
Guoshoujing Telescope (the Large Sky Area Multi-Object Fiber Spectroscopic Telescope LAMOST) is a National Major Scientific Project built by the Chinese Academy of Sciences. Funding for the project has been provided by the National Development and Reform Commission. LAMOST is operated and managed by the National Astronomical Observatories, Chinese Academy of Sciences.

\end{document}

%% file: tab1.tex
\scriptsize
\begin{tabular}{c c c c c c c c c c}
\hline
\hline
\hline
	Age (Gyr)& $N$& $\sigma_{\rm r}$ (km\,s$^{-1}$)& $\sigma_{\phi}$ (km\,s$^{-1}$)& $\sigma_{\rm z}$ (km\,s$^{-1}$)& $\sigma_{\rm r}/\sigma_{\phi}$& $\sigma_{\rm r}/\sigma_{\rm z}$& $\sigma_{\phi}/\sigma_{\rm z}$& $l_{\rm v}$ (deg)& $\alpha$ (deg) \\
\hline
\hline
&&\multicolumn{3}{c}{all stars}\\
\hline
$1.4 \pm 0.3$ & $100$ & $21.7 \pm 2.0$ & $12.5 \pm 1.5$ & $8.2 \pm 1.1$ & $1.7 \pm 0.3$ & $2.7 \pm 0.4$ & $1.5 \pm 0.3$ & $7.3 \pm 7.4$ & $-1.1 \pm 3.8$ \\
$1.9 \pm 0.3$ & $100$ & $21.6 \pm 2.0$ & $17.1 \pm 1.6$ & $10.3 \pm 1.1$ & $1.3 \pm 0.2$ & $2.1 \pm 0.3$ & $1.7 \pm 0.2$ & $0.1 \pm 16.5$ & $9.7 \pm 5.0$ \\
$2.3 \pm 0.2$ & $100$ & $24.7 \pm 2.1$ & $18.9 \pm 1.8$ & $10.6 \pm 1.1$ & $1.3 \pm 0.2$ & $2.3 \pm 0.3$ & $1.8 \pm 0.2$ & $11.3 \pm 12.9$ & $2.7 \pm 3.9$ \\
$2.6 \pm 0.2$ & $115$ & $31.2 \pm 2.4$ & $18.6 \pm 1.7$ & $10.8 \pm 1.1$ & $1.7 \pm 0.2$ & $2.9 \pm 0.4$ & $1.7 \pm 0.2$ & $1.0 \pm 6.3$ & $-3.6 \pm 2.9$ \\
$3.1 \pm 0.2$ & $152$ & $31.9 \pm 2.1$ & $17.5 \pm 1.4$ & $12.3 \pm 1.0$ & $1.8 \pm 0.2$ & $2.6 \pm 0.3$ & $1.4 \pm 0.2$ & $5.3 \pm 4.7$ & $-1.6 \pm 2.8$ \\
$3.6 \pm 0.3$ & $233$ & $32.8 \pm 1.7$ & $18.7 \pm 1.0$ & $13.8 \pm 0.9$ & $1.8 \pm 0.1$ & $2.4 \pm 0.2$ & $1.3 \pm 0.1$ & $6.4 \pm 3.8$ & $-0.7 \pm 2.4$ \\
$4.1 \pm 0.3$ & $391$ & $32.4 \pm 1.3$ & $19.8 \pm 0.8$ & $14.1 \pm 0.7$ & $1.6 \pm 0.1$ & $2.3 \pm 0.1$ & $1.4 \pm 0.1$ & $4.9 \pm 3.5$ & $-2.7 \pm 1.8$ \\
$4.5 \pm 0.3$ & $400$ & $34.9 \pm 1.4$ & $20.5 \pm 0.9$ & $14.1 \pm 0.6$ & $1.7 \pm 0.1$ & $2.5 \pm 0.1$ & $1.5 \pm 0.1$ & $9.3 \pm 2.9$ & $-0.2 \pm 1.6$ \\
$5.0 \pm 0.3$ & $400$ & $37.5 \pm 1.5$ & $22.2 \pm 0.9$ & $16.8 \pm 0.7$ & $1.7 \pm 0.1$ & $2.2 \pm 0.1$ & $1.3 \pm 0.1$ & $9.2 \pm 2.9$ & $2.4 \pm 1.9$ \\
$5.5 \pm 0.4$ & $400$ & $38.4 \pm 1.5$ & $22.8 \pm 1.0$ & $19.5 \pm 0.8$ & $1.7 \pm 0.1$ & $2.0 \pm 0.1$ & $1.2 \pm 0.1$ & $3.3 \pm 2.9$ & $1.5 \pm 2.3$ \\
$6.1 \pm 0.5$ & $361$ & $43.0 \pm 1.8$ & $24.8 \pm 1.1$ & $21.2 \pm 0.9$ & $1.7 \pm 0.1$ & $2.0 \pm 0.1$ & $1.2 \pm 0.1$ & $2.4 \pm 3.0$ & $-3.0 \pm 2.2$ \\
$7.0 \pm 0.5$ & $257$ & $47.5 \pm 2.2$ & $28.9 \pm 1.6$ & $24.5 \pm 1.3$ & $1.6 \pm 0.1$ & $1.9 \pm 0.1$ & $1.2 \pm 0.1$ & $6.9 \pm 3.9$ & $-3.6 \pm 2.8$ \\
$8.0 \pm 0.6$ & $169$ & $50.6 \pm 3.0$ & $32.8 \pm 2.1$ & $29.2 \pm 2.0$ & $1.5 \pm 0.1$ & $1.7 \pm 0.2$ & $1.1 \pm 0.1$ & $11.7 \pm 5.2$ & $-2.3 \pm 4.3$ \\
$9.5 \pm 0.7$ & $157$ & $51.7 \pm 3.3$ & $36.4 \pm 2.4$ & $34.2 \pm 2.2$ & $1.4 \pm 0.1$ & $1.5 \pm 0.1$ & $1.1 \pm 0.1$ & $-5.0 \pm 7.2$ & $-2.2 \pm 5.8$ \\
$10.2 \pm 0.5$ & $100$ & $56.7 \pm 4.3$ & $35.3 \pm 3.0$ & $37.8 \pm 3.0$ & $1.6 \pm 0.2$ & $1.5 \pm 0.2$ & $0.9 \pm 0.1$ & $1.8 \pm 6.6$ & $5.1 \pm 7.3$ \\
\hline
\hline
	&&\multicolumn{3}{c}{${\rm [Fe/H]} > -0.2$\,dex}\\
\hline
$1.4 \pm 0.4$ & $100$ & $21.7 \pm 1.9$ & $12.0 \pm 1.4$ & $8.6 \pm 1.1$ & $1.8 \pm 0.3$ & $2.5 \pm 0.4$ & $1.4 \pm 0.2$ & $7.4 \pm 6.8$ & $-0.6 \pm 3.9$ \\
$1.9 \pm 0.3$ & $100$ & $21.4 \pm 1.9$ & $16.7 \pm 1.6$ & $10.1 \pm 1.1$ & $1.3 \pm 0.2$ & $2.1 \pm 0.3$ & $1.7 \pm 0.2$ & $17.8 \pm 13.1$ & $2.4 \pm 4.8$ \\
$2.4 \pm 0.2$ & $100$ & $27.8 \pm 2.5$ & $18.9 \pm 1.7$ & $10.5 \pm 1.1$ & $1.5 \pm 0.2$ & $2.6 \pm 0.4$ & $1.8 \pm 0.2$ & $-2.8 \pm 9.0$ & $-0.3 \pm 3.2$ \\
$2.8 \pm 0.2$ & $100$ & $32.7 \pm 2.7$ & $18.4 \pm 1.7$ & $11.0 \pm 1.2$ & $1.8 \pm 0.2$ & $3.0 \pm 0.4$ & $1.7 \pm 0.2$ & $0.2 \pm 5.7$ & $0.5 \pm 2.8$ \\
$3.3 \pm 0.2$ & $132$ & $31.3 \pm 2.3$ & $16.8 \pm 1.4$ & $11.9 \pm 1.0$ & $1.9 \pm 0.2$ & $2.6 \pm 0.3$ & $1.4 \pm 0.2$ & $12.6 \pm 4.7$ & $1.7 \pm 2.9$ \\
$3.8 \pm 0.3$ & $196$ & $30.1 \pm 1.6$ & $16.9 \pm 1.1$ & $11.7 \pm 0.8$ & $1.8 \pm 0.1$ & $2.6 \pm 0.2$ & $1.4 \pm 0.1$ & $6.7 \pm 4.0$ & $1.2 \pm 2.4$ \\
$4.3 \pm 0.3$ & $277$ & $34.7 \pm 1.6$ & $17.8 \pm 0.9$ & $12.6 \pm 0.7$ & $2.0 \pm 0.1$ & $2.8 \pm 0.2$ & $1.4 \pm 0.1$ & $9.7 \pm 2.9$ & $-1.7 \pm 1.6$ \\
$4.9 \pm 0.4$ & $304$ & $36.8 \pm 1.7$ & $21.2 \pm 1.0$ & $16.8 \pm 0.8$ & $1.7 \pm 0.1$ & $2.2 \pm 0.1$ & $1.3 \pm 0.1$ & $11.2 \pm 3.1$ & $-1.5 \pm 2.1$ \\
$5.6 \pm 0.4$ & $255$ & $39.3 \pm 2.0$ & $22.1 \pm 1.1$ & $17.7 \pm 0.9$ & $1.8 \pm 0.1$ & $2.2 \pm 0.2$ & $1.3 \pm 0.1$ & $6.0 \pm 3.2$ & $-0.3 \pm 2.4$ \\
$6.4 \pm 0.5$ & $188$ & $42.5 \pm 2.5$ & $23.0 \pm 1.3$ & $18.3 \pm 1.1$ & $1.8 \pm 0.1$ & $2.3 \pm 0.2$ & $1.3 \pm 0.1$ & $4.2 \pm 3.7$ & $-3.0 \pm 2.7$ \\
$7.2 \pm 0.5$ & $106$ & $43.8 \pm 3.3$ & $24.2 \pm 2.0$ & $23.3 \pm 1.7$ & $1.8 \pm 0.2$ & $1.9 \pm 0.2$ & $1.0 \pm 0.1$ & $3.0 \pm 5.0$ & $-3.2 \pm 4.6$ \\
$8.5 \pm 0.8$ & $57$ & $51.8 \pm 5.6$ & $25.8 \pm 2.9$ & $23.3 \pm 2.6$ & $2.0 \pm 0.3$ & $2.2 \pm 0.3$ & $1.1 \pm 0.2$ & $2.0 \pm 5.6$ & $6.2 \pm 4.9$ \\
\hline
\hline
	&&\multicolumn{3}{c}{${\rm [Fe/H]} < -0.2$\,dex}\\
\hline
$3.2 \pm 0.6$ & $100$ & $29.8 \pm 2.4$ & $19.0 \pm 1.9$ & $13.8 \pm 1.3$ & $1.6 \pm 0.2$ & $2.2 \pm 0.3$ & $1.4 \pm 0.2$ & $6.9 \pm 7.8$ & $-7.4 \pm 4.3$ \\
$3.8 \pm 0.3$ & $100$ & $31.3 \pm 2.5$ & $19.7 \pm 1.8$ & $14.8 \pm 1.4$ & $1.6 \pm 0.2$ & $2.1 \pm 0.3$ & $1.3 \pm 0.2$ & $2.0 \pm 7.2$ & $-5.6 \pm 4.3$ \\
$4.5 \pm 0.3$ & $168$ & $35.2 \pm 2.1$ & $21.1 \pm 1.4$ & $16.1 \pm 1.1$ & $1.7 \pm 0.2$ & $2.2 \pm 0.2$ & $1.3 \pm 0.1$ & $1.5 \pm 4.8$ & $3.4 \pm 2.9$ \\
$5.0 \pm 0.4$ & $224$ & $38.2 \pm 2.1$ & $21.0 \pm 1.2$ & $17.3 \pm 1.1$ & $1.8 \pm 0.1$ & $2.2 \pm 0.2$ & $1.2 \pm 0.1$ & $2.3 \pm 3.5$ & $4.1 \pm 2.6$ \\
$5.7 \pm 0.4$ & $196$ & $39.8 \pm 2.4$ & $23.7 \pm 1.4$ & $22.1 \pm 1.3$ & $1.7 \pm 0.1$ & $1.8 \pm 0.2$ & $1.1 \pm 0.1$ & $-1.5 \pm 4.4$ & $2.7 \pm 3.8$ \\
$6.4 \pm 0.5$ & $135$ & $43.6 \pm 3.1$ & $30.0 \pm 2.3$ & $27.9 \pm 2.1$ & $1.5 \pm 0.2$ & $1.6 \pm 0.2$ & $1.1 \pm 0.1$ & $-0.1 \pm 7.8$ & $-4.1 \pm 5.9$ \\
$7.6 \pm 0.6$ & $106$ & $53.4 \pm 4.4$ & $42.5 \pm 4.0$ & $33.9 \pm 2.8$ & $1.3 \pm 0.2$ & $1.6 \pm 0.2$ & $1.3 \pm 0.2$ & $26.8 \pm 9.4$ & $-11.0 \pm 6.6$ \\
$8.9 \pm 0.8$ & $116$ & $56.5 \pm 4.2$ & $37.8 \pm 3.0$ & $35.4 \pm 2.6$ & $1.5 \pm 0.2$ & $1.6 \pm 0.2$ & $1.1 \pm 0.1$ & $3.7 \pm 7.5$ & $-3.7 \pm 6.4$ \\
$10.1 \pm 0.6$ & $110$ & $56.5 \pm 4.1$ & $34.3 \pm 2.5$ & $37.5 \pm 2.8$ & $1.6 \pm 0.2$ & $1.5 \pm 0.2$ & $0.9 \pm 0.1$ & $-0.4 \pm 6.1$ & $6.1 \pm 7.1$ \\
\hline
\hline
\hline
\end{tabular}

%% file: tab2.tex
\scriptsize
\begin{tabular}{c c c c c c c c c c}
\hline
\hline
\hline
	Age (Gyr)& $N$& $\sigma_{\rm r}$ (km\,s$^{-1}$)& $\sigma_{\phi}$ (km\,s$^{-1}$)& $\sigma_{\rm z}$ (km\,s$^{-1}$)& $\sigma_{\rm r}/\sigma_{\phi}$& $\sigma_{\rm r}/\sigma_{\rm z}$& $\sigma_{\phi}/\sigma_{\rm z}$& $l_{\rm v}$ (deg)& $\alpha$ (deg) \\
\hline
\hline
&&\multicolumn{3}{c}{all stars}\\
\hline
$1.6 \pm 0.4$ & $100$ & $26.4 \pm 2.3$ & $12.5 \pm 1.5$ & $10.5 \pm 1.2$ & $2.1 \pm 0.3$ & $2.5 \pm 0.4$ & $1.2 \pm 0.2$ & $-0.4 \pm 5.4$ & $9.0 \pm 4.3$ \\
$2.2 \pm 0.3$ & $100$ & $28.8 \pm 2.6$ & $13.2 \pm 1.5$ & $14.9 \pm 1.4$ & $2.2 \pm 0.3$ & $1.9 \pm 0.3$ & $0.9 \pm 0.1$ & $-3.8 \pm 5.1$ & $13.0 \pm 5.5$ \\
$2.7 \pm 0.3$ & $100$ & $29.2 \pm 2.7$ & $13.7 \pm 1.5$ & $15.7 \pm 1.5$ & $2.1 \pm 0.3$ & $1.9 \pm 0.2$ & $0.9 \pm 0.1$ & $0.8 \pm 4.9$ & $8.9 \pm 6.1$ \\
$3.1 \pm 0.2$ & $121$ & $28.9 \pm 2.4$ & $15.6 \pm 1.5$ & $15.8 \pm 1.3$ & $1.8 \pm 0.2$ & $1.8 \pm 0.2$ & $1.0 \pm 0.1$ & $8.5 \pm 5.0$ & $-3.2 \pm 5.2$ \\
$3.6 \pm 0.3$ & $187$ & $27.1 \pm 1.9$ & $17.7 \pm 1.3$ & $14.5 \pm 1.0$ & $1.5 \pm 0.2$ & $1.9 \pm 0.2$ & $1.2 \pm 0.1$ & $12.2 \pm 5.8$ & $1.2 \pm 4.1$ \\
$4.1 \pm 0.3$ & $280$ & $27.6 \pm 1.5$ & $17.6 \pm 1.0$ & $16.3 \pm 0.8$ & $1.6 \pm 0.1$ & $1.7 \pm 0.1$ & $1.1 \pm 0.1$ & $4.7 \pm 4.7$ & $10.2 \pm 3.8$ \\
$4.7 \pm 0.3$ & $400$ & $30.8 \pm 1.3$ & $17.0 \pm 0.9$ & $17.1 \pm 0.8$ & $1.8 \pm 0.1$ & $1.8 \pm 0.1$ & $1.0 \pm 0.1$ & $-1.4 \pm 2.7$ & $4.3 \pm 2.9$ \\
$5.2 \pm 0.3$ & $400$ & $33.4 \pm 1.5$ & $18.6 \pm 0.9$ & $17.3 \pm 0.8$ & $1.8 \pm 0.1$ & $1.9 \pm 0.1$ & $1.1 \pm 0.1$ & $1.5 \pm 2.8$ & $3.7 \pm 2.4$ \\
$5.7 \pm 0.3$ & $400$ & $33.8 \pm 1.5$ & $20.2 \pm 1.0$ & $20.2 \pm 0.8$ & $1.7 \pm 0.1$ & $1.7 \pm 0.1$ & $1.0 \pm 0.1$ & $3.3 \pm 3.3$ & $4.8 \pm 3.1$ \\
$6.3 \pm 0.4$ & $400$ & $36.6 \pm 1.7$ & $21.9 \pm 1.0$ & $27.0 \pm 1.0$ & $1.7 \pm 0.1$ & $1.4 \pm 0.1$ & $0.8 \pm 0.0$ & $4.6 \pm 3.3$ & $8.8 \pm 5.6$ \\
$7.1 \pm 0.6$ & $365$ & $42.1 \pm 1.9$ & $27.8 \pm 1.4$ & $33.6 \pm 1.4$ & $1.5 \pm 0.1$ & $1.3 \pm 0.1$ & $0.8 \pm 0.1$ & $5.9 \pm 4.4$ & $8.9 \pm 7.3$ \\
$8.0 \pm 0.6$ & $284$ & $45.3 \pm 2.4$ & $31.8 \pm 1.8$ & $35.1 \pm 1.7$ & $1.4 \pm 0.1$ & $1.3 \pm 0.1$ & $0.9 \pm 0.1$ & $4.3 \pm 5.7$ & $-3.0 \pm 8.0$ \\
$9.3 \pm 0.7$ & $187$ & $52.7 \pm 3.4$ & $32.0 \pm 2.2$ & $38.8 \pm 2.3$ & $1.6 \pm 0.2$ & $1.4 \pm 0.1$ & $0.8 \pm 0.1$ & $8.0 \pm 5.0$ & $-0.4 \pm 7.9$ \\
$10.3 \pm 0.4$ & $92$ & $64.1 \pm 5.8$ & $35.0 \pm 3.6$ & $47.0 \pm 4.0$ & $1.8 \pm 0.3$ & $1.4 \pm 0.2$ & $0.7 \pm 0.1$ & $6.5 \pm 6.1$ & $-5.1 \pm 10.9$ \\
\hline
\hline
	&&\multicolumn{3}{c}{${\rm [Fe/H]} > -0.2$\,dex}\\
\hline
$1.6 \pm 0.5$ & $100$ & $26.8 \pm 2.4$ & $10.8 \pm 1.3$ & $11.3 \pm 1.3$ & $2.5 \pm 0.4$ & $2.4 \pm 0.3$ & $1.0 \pm 0.2$ & $-1.2 \pm 4.3$ & $10.5 \pm 4.4$ \\
$2.4 \pm 0.4$ & $100$ & $30.2 \pm 2.7$ & $12.6 \pm 1.4$ & $14.9 \pm 1.4$ & $2.4 \pm 0.3$ & $2.0 \pm 0.3$ & $0.8 \pm 0.1$ & $2.1 \pm 4.4$ & $8.1 \pm 5.2$ \\
$3.0 \pm 0.3$ & $100$ & $28.1 \pm 2.6$ & $15.1 \pm 1.5$ & $14.0 \pm 1.3$ & $1.9 \pm 0.3$ & $2.0 \pm 0.3$ & $1.1 \pm 0.1$ & $9.1 \pm 5.6$ & $3.5 \pm 5.1$ \\
$3.5 \pm 0.3$ & $122$ & $25.9 \pm 2.2$ & $16.4 \pm 1.4$ & $12.2 \pm 1.1$ & $1.6 \pm 0.2$ & $2.1 \pm 0.3$ & $1.3 \pm 0.2$ & $14.2 \pm 6.7$ & $-3.1 \pm 4.1$ \\
$4.1 \pm 0.3$ & $167$ & $27.1 \pm 1.9$ & $16.6 \pm 1.2$ & $15.4 \pm 1.1$ & $1.6 \pm 0.2$ & $1.8 \pm 0.2$ & $1.1 \pm 0.1$ & $13.0 \pm 5.1$ & $7.7 \pm 4.7$ \\
$4.6 \pm 0.3$ & $216$ & $27.4 \pm 1.7$ & $17.3 \pm 1.1$ & $16.2 \pm 1.0$ & $1.6 \pm 0.1$ & $1.7 \pm 0.1$ & $1.1 \pm 0.1$ & $9.4 \pm 4.9$ & $8.5 \pm 4.4$ \\
$5.3 \pm 0.4$ & $247$ & $31.0 \pm 1.7$ & $18.3 \pm 1.1$ & $17.7 \pm 0.9$ & $1.7 \pm 0.1$ & $1.8 \pm 0.1$ & $1.0 \pm 0.1$ & $4.3 \pm 4.1$ & $3.3 \pm 3.7$ \\
$5.9 \pm 0.4$ & $217$ & $34.9 \pm 2.1$ & $19.8 \pm 1.2$ & $20.0 \pm 1.1$ & $1.8 \pm 0.1$ & $1.7 \pm 0.1$ & $1.0 \pm 0.1$ & $-0.1 \pm 3.8$ & $3.0 \pm 3.8$ \\
$6.7 \pm 0.6$ & $131$ & $38.9 \pm 2.9$ & $23.8 \pm 1.9$ & $23.2 \pm 1.7$ & $1.6 \pm 0.2$ & $1.7 \pm 0.2$ & $1.0 \pm 0.1$ & $0.4 \pm 5.5$ & $0.1 \pm 5.5$ \\
$8.0 \pm 0.8$ & $100$ & $41.5 \pm 3.4$ & $28.5 \pm 2.5$ & $24.8 \pm 2.0$ & $1.5 \pm 0.2$ & $1.7 \pm 0.2$ & $1.1 \pm 0.1$ & $12.1 \pm 8.3$ & $3.1 \pm 6.0$ \\
$8.9 \pm 0.6$ & $47$ & $44.7 \pm 5.5$ & $29.6 \pm 3.8$ & $23.2 \pm 3.0$ & $1.5 \pm 0.3$ & $1.9 \pm 0.3$ & $1.3 \pm 0.2$ & $21.6 \pm 9.7$ & $9.2 \pm 7.6$ \\
\hline
\hline
	&&\multicolumn{3}{c}{${\rm [Fe/H]} < -0.2$\,dex}\\
\hline
$3.1 \pm 0.7$ & $100$ & $28.2 \pm 2.7$ & $18.6 \pm 2.0$ & $17.0 \pm 1.6$ & $1.5 \pm 0.2$ & $1.7 \pm 0.2$ & $1.1 \pm 0.2$ & $1.2 \pm 9.0$ & $8.4 \pm 7.1$ \\
$3.9 \pm 0.3$ & $100$ & $29.0 \pm 2.7$ & $19.1 \pm 2.0$ & $19.2 \pm 1.7$ & $1.5 \pm 0.2$ & $1.5 \pm 0.2$ & $1.0 \pm 0.1$ & $3.4 \pm 8.7$ & $5.0 \pm 8.5$ \\
$4.6 \pm 0.3$ & $168$ & $34.3 \pm 2.3$ & $17.1 \pm 1.4$ & $18.1 \pm 1.3$ & $2.0 \pm 0.2$ & $1.9 \pm 0.2$ & $0.9 \pm 0.1$ & $-6.2 \pm 3.8$ & $0.9 \pm 4.1$ \\
$5.2 \pm 0.4$ & $224$ & $35.9 \pm 1.8$ & $18.6 \pm 1.2$ & $18.7 \pm 1.0$ & $1.9 \pm 0.2$ & $1.9 \pm 0.1$ & $1.0 \pm 0.1$ & $-0.7 \pm 3.1$ & $3.4 \pm 3.1$ \\
$5.8 \pm 0.4$ & $196$ & $36.4 \pm 1.8$ & $20.7 \pm 1.1$ & $24.2 \pm 1.1$ & $1.8 \pm 0.1$ & $1.5 \pm 0.1$ & $0.9 \pm 0.1$ & $5.1 \pm 3.5$ & $13.2 \pm 4.6$ \\
$6.7 \pm 0.5$ & $135$ & $41.2 \pm 2.3$ & $26.9 \pm 1.7$ & $32.8 \pm 1.7$ & $1.5 \pm 0.1$ & $1.3 \pm 0.1$ & $0.8 \pm 0.1$ & $6.2 \pm 5.4$ & $8.1 \pm 8.8$ \\
$7.5 \pm 0.6$ & $106$ & $44.5 \pm 2.7$ & $31.9 \pm 2.3$ & $39.5 \pm 2.2$ & $1.4 \pm 0.1$ & $1.1 \pm 0.1$ & $0.8 \pm 0.1$ & $8.0 \pm 7.4$ & $1.2 \pm 18.6$ \\
$8.6 \pm 0.7$ & $116$ & $51.3 \pm 3.6$ & $33.1 \pm 2.7$ & $44.2 \pm 2.7$ & $1.6 \pm 0.2$ & $1.2 \pm 0.1$ & $0.7 \pm 0.1$ & $-1.5 \pm 6.6$ & $11.4 \pm 16.4$ \\
$10.1 \pm 0.6$ & $110$ & $62.8 \pm 4.9$ & $37.4 \pm 3.3$ & $47.0 \pm 3.7$ & $1.7 \pm 0.2$ & $1.3 \pm 0.1$ & $0.8 \pm 0.1$ & $5.6 \pm 6.3$ & $0.1 \pm 10.5$ \\
\hline
\hline
\hline
\end{tabular}